# First principles study of M$_2$InC (M = Zr, Hf and Ta) MAX phases: The effect of M atomic species

F. Sultana[1], M. M. Uddin[1*], M. A. Ali[1], M. M. Hossain[1], S. H. Naqib[2], and A. K. M. A. Islam[2,3]

[1]Department of Physics, Chittagong University of Engineering and Technology, Chittagong-4349, Bangladesh
[2]Department of Physics, University of Rajshahi, Rajshahi 6205, Bangladesh
[3]Department of Electrical and Electronic Engineering, International Islamic University Chittagong, Kumira, Chittagong, 4318, Bangladesh

**Abstract:**

We have studied the physical properties of M$_2$InC (M = Zr, Hf and Ta) MAX phases ternary carbides using density functional theory (DFT) methodology. The structural, elastic and electronic properties are revisited (and found to be in good agreement with recently reported results). The charge density distribution, Fermi surface features, Vickers hardness, dynamical stability, thermodynamics and optical properties have been investigated for the first time. The calculated single crystal elastic constants and phonon dispersion curves endorse the mechanical and dynamical stability of all the compounds under study. The calculated single crystal elastic constants $C_{ij}$ and polycrystalline elastic constants are found to increase with increasing atomic number of M species (M = Zr, Hf and Ta). The values of Pugh ratio and Poisson's ratio revealed the brittleness of the compounds under study associated with strong directional covalent bond with a mixture of ionic contribution. Overlapping of conduction band and valence band at Fermi level notify the metallic nature of M$_2$InC (M = Zr, Hf and Ta) MAX phases. Low values of Vicker hardness indicate the softness of the materials and easy machinability.. The thermodynamic properties, such as the free energy, enthalpy, entropy, specific heat capacity and Debye temperature are evaluated using the phonon dispersion curves and a good correspondence is found with the M atomic species. Electronically important optical properties, e.g., dielectric functions, refractive index, photoconductivity, absorption coefficient, loss function and reflectivity are calculated and discussed in detail in this study.

[*]Corresponding authors: mohi@cuet.ac.bd

*Keywords:* MAX phase, $M_2InC$ (M =Zr, Hf, Ta) compounds, Physical properties, First principles study, Phonon dispersion curve

## 1. Introduction

The term "$M_{n+1}AX_n$ phases" was coined by Barsoum in 2000 for the first time [1]. In the general formula $M_{n+1}AX_n$ with n = 1-3, M = early transition metal; A= which? group element and X= C and/or N). The MAX phases $M_2AX$, $M_3AX_2$ and $M_4AX_3$ are referred as 211, 312, and 413 subject to the value of *n*. The $M_{n+1}AX_n$ phases are crystallized in the hexagonal structure belonging to the space group of *$P6_3/mmc$*. The MAX phase nano layered ternary compounds are suitable for many technological applications owing to an unique combination of both metallic and ceramic properties. Like metals, they exhibit good electrical and thermal conductivity, machinability, low hardness, thermal shock resistance, and damage tolerance. On the other hand, these compounds possess ceramiclike high elastic moduli, high melting temperature, and oxidation and corrosion resistance [2-17]. There are forty eight 211 phases that have already been listed without considering possible solid solutions, however more have been theoretically predicted [1,18-22]. Furthermore, very recently the MAX phase materials are used as a precursor to synthesize atomically thin two-dimensional materials with many attractive physical features, the so called MXenes [23].

The extensive research effort has been paid on both theoretical and experimental study of $M_2AX$ phases [2-17], but a few of them fully studied yet. Among the compounds, there are few studies on technologically important $M_2InC$ (M = Zr, Hf and Ta) MAX phases [10, 24-29]. The phases $Zr_2InC$ (ZIC) and $Hf_2InC$ (HIC) have already been synthesized; however the phase Ta$_2$InC (TIC) is neither synthesized nor studied in detail yet.

The ternary carbide ZIC has been synthesized and oxidation behavior with structural parameters are reported by S. Gupta et al. [24]. Manoun et al. [25] have also synthesized ZIC and measured compression under pressure. It is reported that the phase ZIC oxidizes readily to form In$_2$O$_3$ and transition metal oxide and it cannot be used for extended time in air. Electrical and thermal properties such as heat capacities, thermal expansion coefficients, thermal and electrical conductivities of the ternary compound HIC have also been studied [26]. Medkour et al. [27] have investigated the structural and electronic properties of ZIC and HIC. The structural and elastic properties of ZIC and HIC have been studied by He et al. [28]. The elastic properties of ZIC, HIC and TIC have also been investigated by A. Bouhemadou [29].

In order to recommend a compound especially for technological applications, detail theoretical investigations on the physical properties of the material are necessary. Study of dynamical stability of the materials is important for practical application under extreme pressure and temperature conditions. Moreover, the thermodynamic properties provide the important supplementary information regarding the behavior of materials under high pressures and temperatures, which are considered as the basis of many industrial applications [30]. The optical properties are directly related to the electronic properties of materials which exhibits the electronic response of the materials subjected to radiation. To select a material for optoelectronic devices, the information about absorption coefficient and refractive index of the materials is necessary [31]. Furthermore, study of reflectivity of MAX phases is used to predict the suitability of materials as coating materials to reduce solar heating [32]. Therefore, study of these physical attributes of M$_2$InC (M = Zr, Hf and Ta) MAX phases is desirable from research as well application point of view.

Therefore, we are motivated to study the dynamical stability, thermodynamic and optical properties of $M_2InC$ (M = Zr, Hf and Ta) MAX phases for the first time in detail. Moreover, the structural, elastic and electronic properties are revisited with some new additional information on charge density mapping, Fermi surface topology, Mulliken analysis, and Vickers hardness. The results are discussed on the basis of the electronic configuration of the M species (M = Zr, Hf and Ta).

## 2. Computational methodology

The Cambridge serial total energy package (CASTEP) code [33] is used for the first-principles quantum mechanical calculations wherein the pseudo-potential plane-waves (PP-PW) approach based on the density functional theory (DFT) [34] are employed. The Generalized Gradient Approximation (GGA) method with default the Perdew-Burke-Ernzerhof (PBE) formalism is adopted as the exchange and correlation terms [35]. The interactions between electrons-ions are represented by pseudopotentials within Vanderbilt-type ultrasoft formulation [36]. The following parameters are used for the calculations: plane wave cut-off energy is set to 500 eV for all calculations to ensure convergence, the Monkhorst-Pack scheme [37] is used for $k$-points (9x9x2) sampling integration set to ultrafine quality over the first Brillouin zone for the crystal structure optimization, the tolerances for self-consistent field is $5.0\times10^{-7}$ eV/atom, energy is $5.0\times10^{-6}$ eV/atom, maximum force is 0.01 eV/Å, maximum displacement is $5.0\times10^{-4}$ Å, and a maximum stress of 0.02 GPa. The electronic wave functions and consequent charge density as well as the structural parameters of hexagonal $Zr_2InC$, $Hf_2InC$ and $Ta_2InC$ are calculatedfollowing the Broyden–Fletcher–Goldfarb-Shenno (BFGS) [38] minimization technique. The total energy of each cell is calculated by the periodic boundary conditions. Elastic constants are calculated by the 'stress-strain' method in-built in the CASTEP program. The calculated elastic constant tensors $C_{ij}$ are used to evaluate bulk modulus $B$ and shear modulus $G$.

## 3. Results and discussion

### 3.1 *Structural properties*

As mentioned earlier the MAX phases $M_2InC$ (M = Zr, Hf and Ta) crystallizes in the *P6$_3$/mmc* (194) space group with hexagonal structure. The unit cell structure of $M_2InC$ (M = Zr, Hf and Ta) compounds is depicted in Fig.1. The unit cell contains two molecules and 8 atoms with the Wyckoff positions of the four Zr/Hf/Ta atoms are located at (1/3, 2/3, $z_M$), C atoms are located at the positions (0, 0, 0), the In atoms are located at (1/3, 2/3, 3/4) [39] as shown in Fig. 1. Our calculated results are compared with other reported results as listed in Table 1, which testifies the reliability and accuracy of our calculations. The lattice constants *a* and *c* of the $Zr_2InC$ are of 1.34 % and 0.3% larger than that of experimental values [25] and the lattice constants *a* and *c* of the $Hf_2InC$ are 1.54% & 1.06% larger than experimental values [26] but are found in very good agreement with the earlier theoretical values [27-29]. However, experimental results for the $Ta_2InC$ are not available, but the lattice constants *a* and *c* are of 2.92% and 3.18% higher than the reported values [29].

**Table 1** Calculated lattice parameters (*a* and *c*, in Å), hexagonal ratio *c/a*, internal parameter $z_M$, unit cell volume *V* (Å$^3$) for the MAX phases $Hf_2Inc$, $Zr_2InC$ and $Ta_2InC$.

| Phases | *a* | *c* | *c/a* | $z_M$ | V | Ref. |
|---|---|---|---|---|---|---|
| **Zr$_2$InC** | 3.36 | 15.11 | 4.49 | 0.0825 | 148.036 | This work |
| | 3.349 | 14.91 | 4.45 | - | | Expt. [25] |
| | 3.358 | 15.09 | | 0.08123 | 147.437 | Theo. [27] |
| | 3.35 | 15.04 | 4.49 | 0.0817 | 146.62 | Theo. [28] |
| | 3.307 | 14.806 | 4.477 | 0.0825 | - | Theo. [29] |
| **Hf$_2$InC** | 3.36 | 14.88 | 4.42 | 0.0890 | 145.781 | This work |
| | 3.309 | 14.723 | 4.47 | - | 138.919 | Expt. [26] |
| | 3.351 | 14.803 | | 0.0836 | 143.72 | Theo. [27] |
| | 3.35 | 14.79 | 4.41 | 0.0836 | 143.72 | Theo. [28] |
| | 3.304 | 14.553 | 4.404 | 0.0844 | | Theo. [29] |
| **Ta$_2$InC** | 3.21 | 14.57 | 4.53 | 0.0843 | 130.416 | This work |

|  | 3.119 | 14.120 | 4.527 | 0.0843 |  | Theo. [29] |

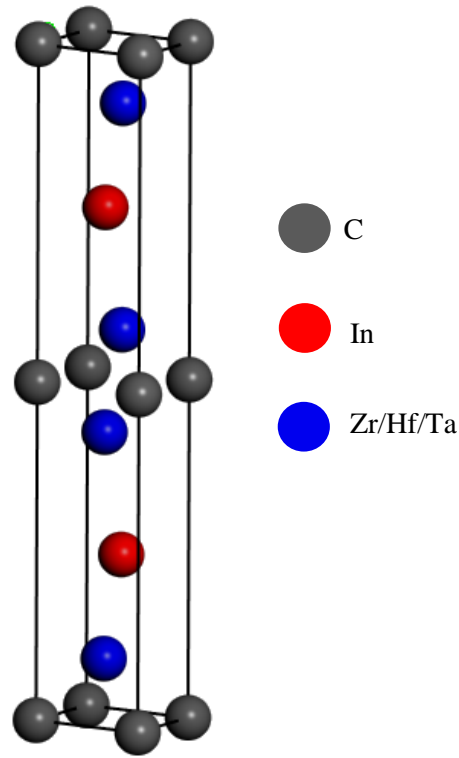

**Fig. 1:** The unit cell of the $M_2InC$ (M = Zr, Hf and Ta) compounds.

3.2 *Mechanical properties*

The mechanical stability, stiffness, brittleness, ductility, and elastic anisotropy of a material can be obtained from the elastic constants which are important to select a material for engineering applications. The five independent single crystal elastic constants $C_{ij}$ (since $C_{66} = (C_{11} - C_{12})/2$) and polycrystalline elastic moduli are given in Table 2. The compounds are mechanically stable since the single crystal elastic constants completely satisfy the Born [40] criteria: $C_{11} > 0$, $C_{11}-C_{12} > 0$, $C_{44} > 0$, $(C_{11} + C_{12}) C_{33} - 2C_{13} > 0$ for a hexagonal system. For all compounds $C_{11} > C_{33}$ (Table 2), hence, the atomic bonding is stronger along *a*-axis compared to that of along *c*-axis. Since $C_{11}$ and $C_{33}$ are larger than $C_{44}$ the linear compression along the crystallographic *a*- and *c*-axis is difficult compared to the shear deformation. From Table 2, it is seen that the value $C_{11}$ is smaller for $Zr_2InC$ (279.3 GPa) than those of $Hf_2InC$ (331.6

GPa) and Ta$_2$InC (451.7 GPa). Therefore Zr$_2$InC has a lower resistance against the principal strain $\varepsilon_{11}$. Moreover, Zr$_2$InC has also smaller value of $C_{44}$ of 94.6 GPa than those of Hf$_2$InC (101.7 GPa) and Ta$_2$InC (161.2 GPa) also indicating smaller resistance to basal shear deformation. Our computed elastic constants for M$_2$InC (M = Zr, Hf and Ta) are found in good agreement with reported theoretical results [29]. The elastic constants $C_{ij}$ are found to increased with increasing atomic number of M (M = Zr, Hf and Ta).

**Table 2.** The calculated elastic constants, $C_{ij}$ (GPa), bulk modulus, $B$ (GPa), shear modulus, $G$ (GPa), Young's modulus, $Y$ (GPa), Pugh ratio, $G/B$, and Poisson ratio, $v$ and Cauchy Pressure of M$_2$InC (M = Zr, Hf and Ta) MAX phases.

| Phases | $C_{11}$ | $C_{12}$ | $C_{13}$ | $C_{33}$ | $C_{44}$ | $B$ | $G$ | $Y$ | $G/B$ | $v$ | Cauchy Pressure | Ref |
|---|---|---|---|---|---|---|---|---|---|---|---|---|
| Zr$_2$InC | 279 | 66 | 75 | 255 | 94 | 137 | 99 | 239 | 0.72 | 0.208 | -28.5 | This work |
| | 286 | 62 | 71 | 248 | 83 | 136 | 95 | 232 | 0.70 | 0.215 | | Theo.[29] |
| Hf$_2$InC | 331 | 87 | 90 | 284 | 101 | 168 | 109 | 270 | 0.64 | 0.232 | -14.8 | This work |
| | 309 | 81 | 80 | 273 | 98 | 152 | 105 | 256 | 0.69 | 0.219 | | Theo.[29] |
| Ta$_2$InC | 452 | 147 | 197 | 425 | 161 | 248 | 157 | 390 | 0.63 | 0.23 | -14.02 | This work |
| | 396 | 102 | 133 | 345 | 133 | 208 | 133 | 329 | 0.63 | 0.236 | | Theo.[29] |

The polycrystalline elastic moduli ($B$, $G$, $Y$, and $v$) have been calculated from the single crystal elastic constants through the Voigt-Reuss-Hill formula [41-42] as shown in Table 2. In addition, the Young's modulus ($Y$) and Poisson's ratio ($v$) can also be obtained using well-known relationships [43, 44]. The average bond strength of constituent atoms for a given compound can be understood by the bulk modulus [45]. The calculated value of $B$ indicates moderately strong average bonding strength of the atoms involved in these compounds. The values of B (G) for Zr$_2$InC, Hf$_2$InC and Ta$_2$InC are 137 (99) GPa, 168 (109) GPa and 248 (157) GPa, respectively. Although, the bulk modulus and shear modulus do not measure the hardness of solids but the values are comparatively greater for harder one. Therefore, it is expected that Ta$_2$InC should be harder compared to the other compounds. It is also found from Table 2 that Ta$_2$InC is stiffer than Zr$_2$InC and Hf$_2$InC because of the higher value of Young's modulus ($Y$). The value of $G/B$ (Pugh ratio [46]) is found to be greater than 0.57 and the Poisson's ratio $v$ (are 0.208, 0.232 and 0.23 for $Zr_2InC$, $Hf_2InC$ and $Ta_2InC$, respectively) is less than Frantsevich's criterion (0.26) [47], indicating the brittle nature of the phases studied here. Moreover, the solids with only covalent bond have value of $v$ is around 0.1 while the ionic solids have the value of $v$ of about 0.25 [48]. The values of $v$ are 0.20, 0.23 and 0.23 for $Zr_2InC$, $Hf_2InC$ and $Ta_2InC$, respectively, indicating to the existence of mixture of covalent and ionic bonding in these compounds. The Cauchy pressure ($C_{12} - C_{44}$) [49] also provides the bonding nature of solids. It is positive for solids in which metallic bonding is dominating while it is negative for solids in which strong covalent bonding dominates. The dominant bonding within the compounds considered here is covalent in nature. The polycrystalline elastic modulii are also increased with atomic number of M as noted from singe crystal elastic constants.

*3.3 Electronic properties*

*3.3.1 Band structure and density of states (DOS)*

Since the information regarding electrical properties can be known from electronic band structure therefore, we have calculated band structure and density of state (DOS) of $M_2InC$ (M = Zr, Hf and Ta) MAX phases. The calculated energy band structure of $M_2InC$ (M = Zr, Hf and Ta) along high-symmetry $k$ points in the first Brillouin zone using equilibrium lattice parameters is shown in Figs. 2(a-c). The band structures indicate the metallic nature due to noticeable overlapping of conduction bands and valence bands at the Fermi level (black dashed line). Moreover, the anisotropy in electrical conductivity is clearly observed due to lower $c$ axis energy dispersion. It is seen from band structure that the electrical conductivity along the $c$ direction should be lower than that in the $ab$ plane.

The natures of chemical bonding and structural features of the compound can be understood using total and partial density of states (DOS). The calculated total and partial DOS of the

M$_2$InC phases are shown in Figs. 2(d-f), where the vertical dashed line denotes the Fermi level, $E_F$. The valence and conduction bands are overlapped considerably thereby no band gap is at the $E_F$. The total energy scale is chosen from -12 to 6 eV. The value of TDOS at $E_F$ is found to be 2.4, 2.25 and 3.1 states per eV for the phases M$_2$InC (M = Zr, Hf and Ta), respectively. The Zr-4$d$, Hf-5$d$ and Ta-5$d$ electrons are mainly contributing in the electronic conduction properties making by far the largest contribution to the TDOS at the $E_F$ (Figs. 2(d-f)). However, In and C atoms are not significantly associated with the metal-like conduction since they contribute weakly to the TDOS at the $E_F$.

The valence band can be resolved into three sub-bands. The lowest energy sub-band (-8 eV to -10 eV) is derived mainly from C-2$s$ states. The middle sub-band (-8 eV to -2 eV) of broad nature is formed primarily from the contributions due to C-2$p$ and In-5$s$ orbitals. The top sub-band crossing the Fermi level is derived mainly from Zr-4$d$, Hf-5$d$ and Ta-5$d$ electronic states as shown in Figs. 2(d-f). The strong covalent bonding is formed between Zr/Hf/Ta-C ($pd$) since they are mixed together strongly. There is another covalent bond formed at the higher energy states between Zr/Hf/Ta-In ($pd$) states but comparatively weaker than the previous one. It is noted that the hybridization between Zr/Hf/Ta-C ($pd$) states is occurred at lower energy for Ta$_2$InC and higher energy for Zr$_2$InC which may result in the larger elastic modulii and higher Vickers hardness for Ta$_2$InC compared to the other two phases. This can also be observed from the charge density distribution in the following section.

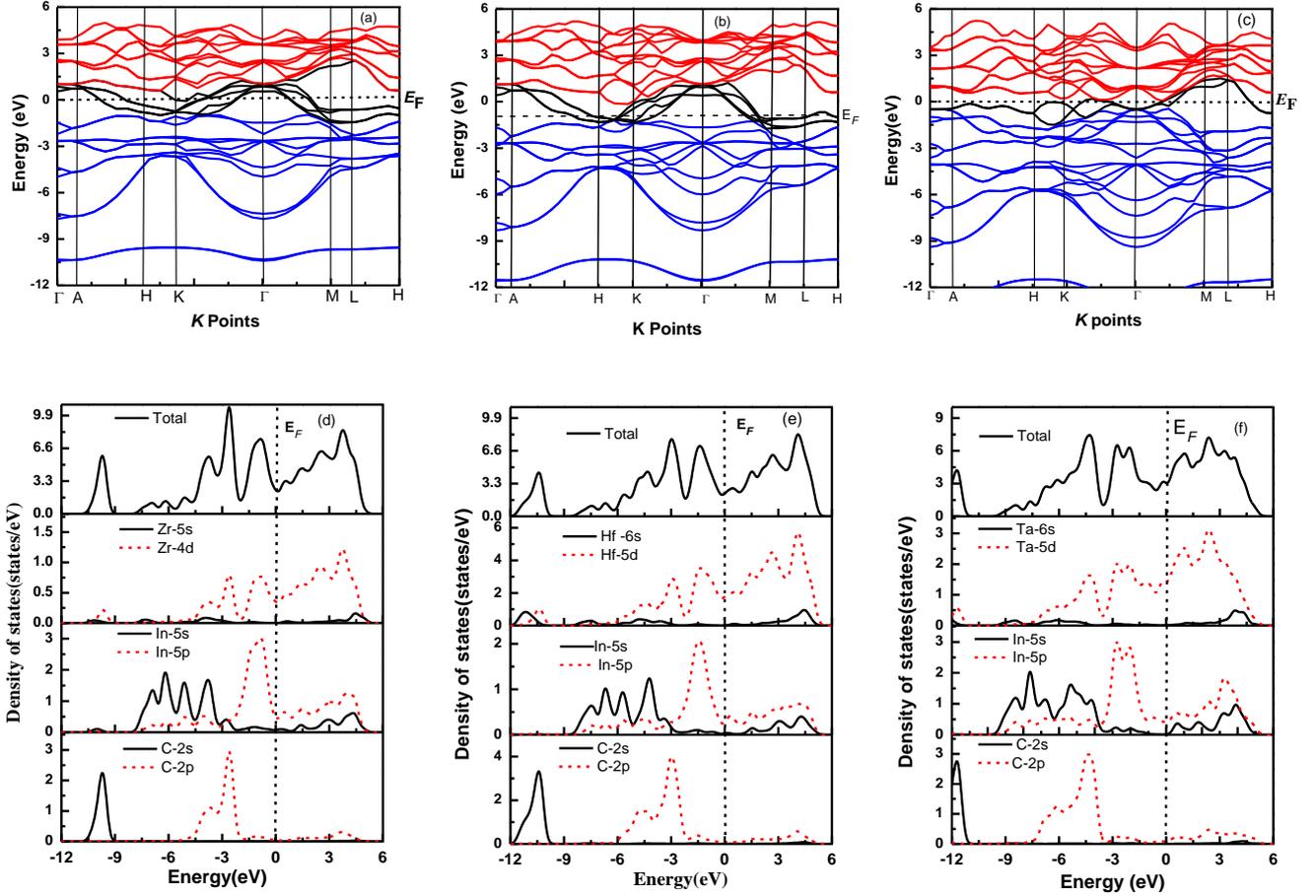

**Fig. 2.** Electronic band structures of the compounds $Zr_2InC$ (a), $Hf_2InC$ (b) and $Ta_2InC$ (c). The total and partial density of states (DOS) $Zr_2InC$ (d), $Hf_2InC$ (e) and $Ta_2InC$ (f).

### 3.3.2 Charge density mapping and Fermi surface

Electronic charge density distribution mapping in contour form (in the units of e/Å$^3$) along the (101) crystallographic plane has been calculated to uncover the nature of chemical bonding in the MAX phases of $M_2InC$ as shown in Figs. 3(a-c). The electron densities due to different chemical bonds in the compounds are the main feature of this mapping. The regions of positive or negative electrons density can be understood by the accumulation and depletion of electronic charges, respectively. The covalent bonds arise from the preferential accumulation of charges between two atoms whereas the balancing of positive or negative charge at the atomic position indicates the ionic bonding [50]. The strong charge accumulation regions occur at the position of C atoms and Zr, Hf and Ta atoms as shown in

Fig. 3(a-c), consequently robust covalent bonding is formed between C-Zr, C-Hf and C-Ta atoms. (Fig. 3(a)). Another covalent bonding between In and Zr/Hf/Ta atoms is also observed but this is comparatively weaker than the previous one (Fig. 3 (a-c)). Furthermore, it is also seen that there is signs of charge balance around Zr/Hf/Ta with the carbon atoms, exhibiting a small degree of ionic bonding. Therefore, the bonding in $M_2InC$ (M = Zr, Hf and Ta) is expected to be a mixture of covalent andionic. It is also noted here that the accumulation charge at Zr/Hf/Ta position follows the sequence: $Zr_2InC$ < $Hf_2InC$ < $Ta_2InC$; results the stronger covalent bonding between Ta-C (*pd*) atoms, which is in good agreement with analysis in the previous section.

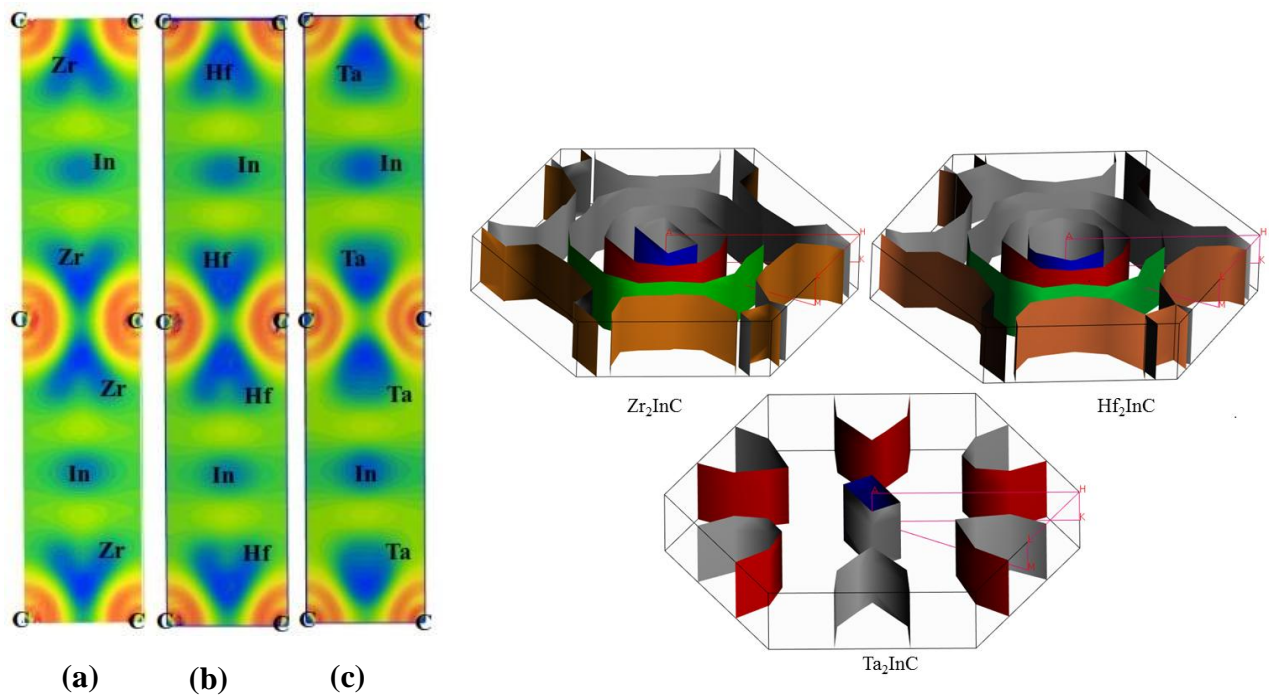

Fig. 3. Electronic charge density mapping (show the scale) of $Zr_2InC$ (a), $Hf_2InC$ (b) and $Ta_2InC$ (c) phases. The Fermi surfaces of $M_2InC$ (M = Zr, Hf and Ta) MAX phases (right).

The Fermi surface topology of the MAX phases $M_2InC$ (M = Zr, Hf and Ta) has been explored in the equilibrium structure at zero pressure as shown in Fig. 3 (right). The calculated Fermi surfaces show both electron and hole-like sheets. The shape of the Fermi surfaces for the ZIC and HIC phases are quite similar, however it is somewhat different for the TIC. It is seen that the central sheet (inner blue) is of cylindrical shape with hexagonal cross section and centered along the Γ-A direction of the Brillouin zone for the three phases (Fig. 3(right)). Second sheet of the ZIC and HIC phases (red color) is also of cylindrical shape, much closer and surrounding the first sheet while for the TIC, it has six tuning fork like shape in the six corners far from the first sheet. The third sheet is hexagonal cylinder with six little curved plane along M–L directions for the ZIC and HIC phases. Six up curved ribbon type (red accent darker color) tubes are in each corner of fourth sheet for the ZIC and HIC phases as shown in Fig. 3(right). The Fermi surface of the ZIC and HIC phases is due to the low-dispersive of Zr5$d$, Hf 5$d$ and In- 5$p$ states, respectively, on the other hand, for TIC it is due to high-dispersive Ta 5$d$ and In 5$p$ orbitals as is assured from DOS seen in Fig. 2 (d-f).

*3.3.3 Mulliken atomic and bond overlap populations*

The information regarding nature of chemical bonding in the compounds can be obtained from the Mulliken atomic population (MAP), effective valence charge (EVC) and bond overlap population (BOP) analyses [51]. The difference between the formal ionic charge and Mulliken charge on the anion species within a crystal is represented by the EVC. If the value of EVC is zero then the chemical bond exist as an ideal ionic bond in the compounds, however the positive EVC indicates the presence of covalent bond and the high values signifies the high level of covalency in chemical bonds. The calculated EVC of $M_2InC$ phases is depicted in Table 3.

The band overlap population (BOP) of the $M_2InC$ (M = Zr, Hf and Ta) MAX phases are also presented in Table 4. The high value of BOP represents the strong covalent bond while the low values represent the ionicity of chemical bonding. For instance, the C-Zr, C-Hf, and C-Ta bonding possess stronger covalent bonding than In-Zr, In-Hf and In-Ta bonding in $M_2InC$ (M = Zr, Hf and Ta). In the case of $Ta_2InC$, C-Ta bonding is comparatively stronger than Zr-In, In-Hf, and In-Ta bonds in the $M_2InC$ compounds. The overall bond strength is greater in $Hf_2InC$ and $Ta_2InC$, therefore, the hardness value is expected higher for them.

**Table 3.** Mulliken atomic and bond overlap population of the $M_2InC$ compounds.

| | | Mulliken atomic population | | | | | | Mulliken bond overlap population | | | |
|---|---|---|---|---|---|---|---|---|---|---|---|
| **Phases** | Atoms | $s$ | $p$ | $d$ | Total | Charge (e) | EVC (e) | Bond | Bond number $n^\mu$ | Bond length $d^\mu$ (Å) | Bond overlap population $P^\mu$ |
| $Zr_2InC$ | C | 1.49 | 3.31 | 0.00 | 4.80 | -0.80 | 4.80 | C-Zr | 4 | 2.30 | 1.07 |
| | Zr | 2.26 | 6.58 | 2.68 | 11.53 | 0.47 | 11.53 | Zr-In | 4 | 3.198 | 0.11 |
| | In | 1.28 | 1.90 | 9.98 | 13.15 | -0.15 | 13.15 | | | | |
| $Hf_2InC$ | C | 1.55 | 3.33 | 0.00 | 4.88 | -0.88 | 4.88 | C-Hf | 4 | 2.3 | 1.41 |
| | Hf | 0.48 | 0.36 | 2.78 | 3.62 | 0.38 | 3.62 | In-Hf | 4 | 3.14 | 0.25 |
| | In | 0.93 | 1.97 | 9.97 | 12.87 | 0.13 | 12.87 | In-Hf | 4 | 4.96 | 0.05 |
| $Ta_2InC$ | C | 1.50 | 3.22 | 0.00 | 4.72 | -0.72 | 4.72 | C-Ta | 4 | 2.22 | 1.37 |
| | In | 0.79 | 1.88 | 9.97 | 12.64 | 0.36 | 12.64 | In-Ta | 4 | 4.87 | 0.04 |
| | Ta | 0.48 | 0.50 | 3.84 | 4.82 | 0.18 | 4.82 | | | | |

*3.4 Vickers Hardness*

We have already discussed about bulk and shear modulus of $M_2InC$ (M = Zr, Hf and Ta). But the intrinsic hardness is different from bulk modulus or shear modulus [52]. The hardness of material plays an important role in its applications. From this point of view we have also calculated using established formalism [53, 54] which is fundamental criteria for an engineering point of view as the resistance to wear by either friction or erosion that increases with hardness [15]. The calculated Vickers hardness values are found to be 1.05, 3.45, and 4.12 GPa (Table 5) for the phases ZIC, HIC and TIC, respectively. It is seen again that the value of Vickers hardness is largest for TIC which is good agreement with elastic modulii, partial density of states and charge density mapping. The values are typical for the MAX phase compounds and indicating the softness of the phases which are easily machinable.

**Table 5. (Should be Table 4)** Calculated Muliken bond overlap population of μ-type bond $P^{\mu}$, bond length $d^{\mu}$, metallic population $P^{\mu'}$, bond volume $v_b^{\mu}$, Vickers hardness of μ-type bond $H_V^{\mu}$ and $H_V$ of M$_2$InC (M = Zr, Hf and Ta) compound.

| Phases | Bond | $d^{\mu}$ | $P^{\mu}$ | $P^{\mu'}$ | $v_b^{\mu}$ | $H_V^{\mu}$ | $H_V$ |
|---|---|---|---|---|---|---|---|
| **Zr$_2$InC** | C-Zr | 2.30 | 1.07 | 0.0709 | 10.0519 | 15.79 | 1.05 |
| | In-Zr | 2.30 | 1.07 | 0.7091 | 37.009 | 0.07 | |
| **Hf$_2$InC** | C-Hf | 2.30 | 1.41 | 0.0757 | 2.6927 | 189.46 | 3.45 |
| | In-Hf | 3.14 | 0.25 | 0.0757 | 7.4245 | 4.57 | |
| | In-Hf | 4.96 | 0.05 | 0.0757 | 36.4453 | -0.05 | |
| **Ta$_2$InC** | C-Ta | 2.22 | 1.37 | 0.0858 | 2.8402 | 166.84 | 4.12 |
| | In-Ta | 2.22 | 0.04 | 0.0858 | 32.604 | -0.10 | |

*3.5 Phonon dispersion curve*

Dynamical stability of material and vibrational contribution in the various thermodynamic properties such as thermal expansion, Helmholtz free energy, and heat capacity can be understood by the phonon dispersion curve (PDC) along with phonon density of states (PHDOS) [55, 56]. The PDC and PDOHS of the phases ZIC, HIC and TIC along the high symmetry direction of the crystal Brillouin zone (BZ) have been calculated using the density functional perturbation theory (DFPT) linear-response method [57] and are shown in Fig.4. The optical behavior strongly depends on the optical branches that are situated at the top in the phonon dispersive curve shown in Fig. 4. The values of transverse optical (TO) and longitudinal optical (LO) phonon modes corresponding frequencies at the zone center (Γ) are found to be 17.4, 16.1 THz (ZIC), 19.2, 17.6 THz (HIC) and 19.6, 21.1 THz (TIC), respectively.

The phonon density of states (PHDOS) of ZIC, HIC and TIC phases have been calculated and illustrated in Figs. 4(b, d, f), respectively, along with PDC curves to distinguish the bands by corresponding peaks. It is observed that the flatness of the bands for TO produces prominent peaks for the phases as shown in Fig. 4 (b, d, f). The separation between top of the LO and bottom of the TO modes at the Γ point are found to be 1.3, 1.6 and 1.5 THz for the phases ZIC, HIC and TIC, respectively. The PDC curves of MAX phases under consideration show positive phonon frequencies depicted in Fig. 4 (b, d, f) No negative frequencies exist indicating the phases are dynamically stable. Moreover, the elastic constants of the phases are also supported the condition for mechanical stability (Table 2).

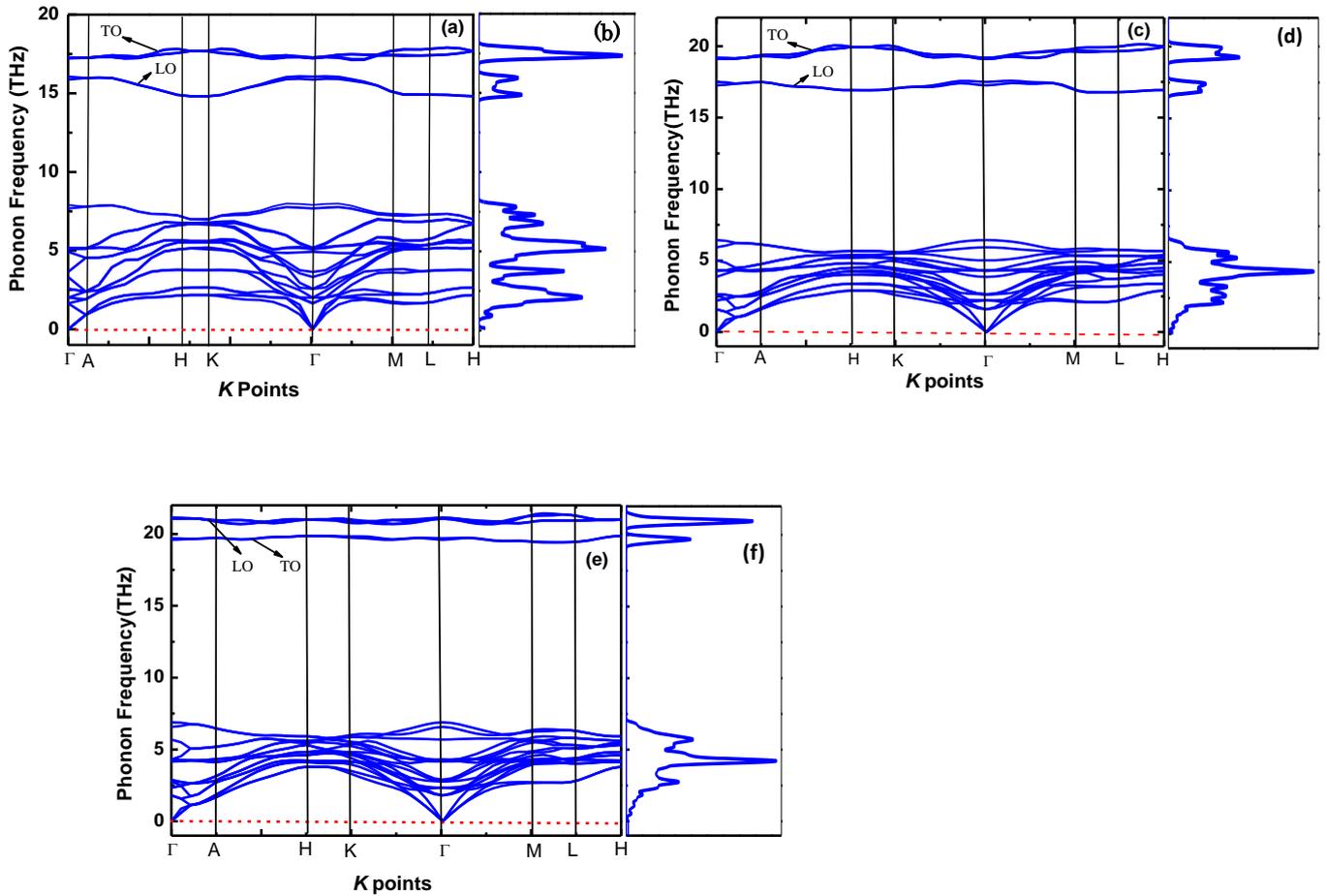

**Fig. 4.** Phonon dispersion curve (PDC) and phonon density of states (PHDOS) of $Zr_2InC$ (a, b), $Hf_2InC$ (c, d) and $Ta_2InC$ (e, f), respectively. The dashed line (red) is at zero phonon frequency. No imaginary (negative) frequency exists in the three phases.

*3.6 Thermodynamical properties*(change font size)

The PHDOS are used to calculate different temperature-dependent thermodynamical potential functions such as Helmholtz free energy $F$, Enthalpy $E$, entropy $S$, phonon specific heat $C_v$ and the Debye temperature $\Theta_D$ at zero pressure using quasi-harmonic approximation

[58, 59] as shown in Fig. 5. The potential functions are calculated in the temperature range 0-1000 K where no phase transitions are expected. It can be seen from Fig. 5 (a) for the phases, below 100 K, the values of $E$, $F$ and $TS$ are almost zero. A nonlinear decrease of $F$ is observed above 100 K, which is very common scenario and it becomes more negative during the course of any natural process. The thermal disturbance enhances the disorder with increasing temperature in the system results in entropy increase as shown in Fig. 5 (a). As expected for solids, an increasing trend of enthalpy ($E$) with the increase of temperature is observed.

Fig.5 (b) illustrates constant volume specific heat $C_v$ of the phases as a function of temperature. It is seen that curves for ZIC and HIC are almost identical; however it is different for TIC phase. It follow common rend that at very low temperatures, the $C_v$ of the phases is strongly dependent on the temperature and obeys the Debye model which is proportional to $T^3$ [60] during the phonon coupling process of crystal lattice vibration. Nevertheless, when the temperature is higher, $C_v$ does not depend strongly on the temperature and the the classical Dulong-Petit law is recovered [61].

The temperature dependence of $\Theta_D$ for the phases has been calculated using PHDOS shown in Fig. 5 (c). It is seen that the $\Theta_D$ increases with increasing temperature indicating the change of the vibration frequency of particles under temperature effects. The lower the value of $\Theta_D$ indicates the weaker bonds in solids, the heat capacity reach the classical Dulong–Petit value at lower temperature. From Fig. 5(c) it is clearly observed that the value of $\Theta_D$ at room temperature is lowest for ZIC and highest for TIC which is in good agreement with our previous results.

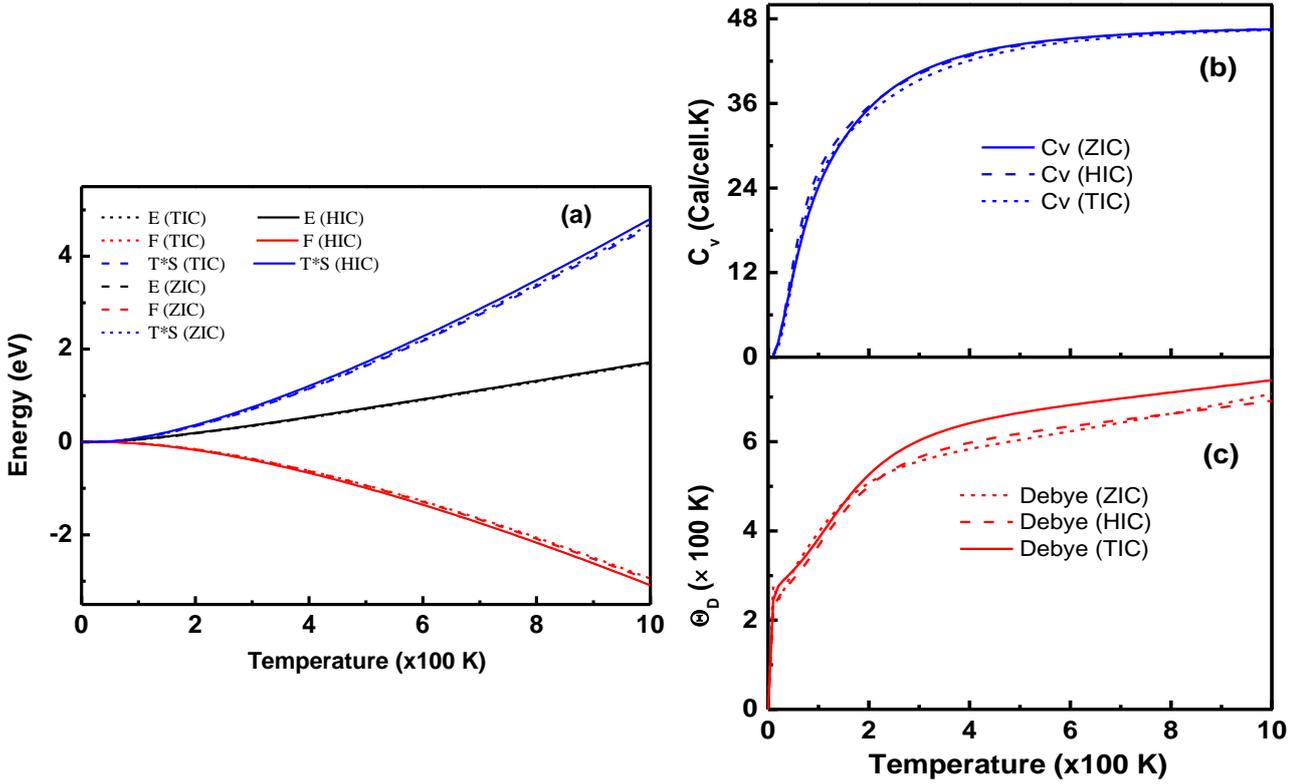

**Fig. 5.** Temperature dependence of the thermodynamical potential functions, $C_v$ and $\Theta_D$ of $Zr_2InC$ (ZIC), $Hf_2InC$ (HIC) and $Ta_2InC$ (TIC) phases.

*3.7 Optical properties*

The optical properties of MAX phases $M_2InC$ (M = Zr, Hf, Ta) have been calculated for photon energies up to 20 eV for polarization vectors [100] and [001] for the first time. It is seen that the curves for two polarization directions are similar in nature and almost identical, therefore data for [100] direction is illustrated in Figs. 6 (a-h). The compounds under study are metallic in nature (Fig. 2); a Drude term with unscreened plasma frequency of 3 eV and damping of 0.05 eV has been used. A Gaussian smearing of 0.5 eV is used for all the calculations. This smears out the Fermi level, so that *k*-points will be more effective on the Fermi surface. The optical properties of MAX phases ZIC, HIC and TIC are determined by the frequency-dependent dielectric function $\varepsilon(\omega) = \varepsilon_1(\omega) + i\varepsilon_2(\omega)$. The value of $\varepsilon_2(\omega)$ can be

calculated from the momentum matrix elements between the occupied and unoccupied electronic states as [62]

$$\varepsilon_2(\omega) = \frac{2e^2\pi}{\Omega\varepsilon_0} \sum_{k,v,c} |\psi_k^c|u.r|\psi_k^v|^2 \delta(E_k^c - E_k^v - E)$$

Here $\omega$ is the light frequency, $u$ is the vector defining the polarization of the incident electric field, $e$ is the electronic charge. $\psi_k^c$ and $\psi_k^v$ are the conduction and valence band wave functions at $k$, respectively. From this imaginary part, the real part can be extracted via the Kramers-Kronig equation. Once these are known, all the optical parameters can be calculated.

The real part $\varepsilon_1(\omega)$ and imaginary part of $\varepsilon_2(\omega)$ of the dielectric function of the phases are depicted in Figs. 6 (a, b). The peaks in $\varepsilon_2(\omega)$ are associated with the electron excitations. The spectra at the low energy region (infrared region) arise due to the intra-band transition of electrons. The peaks are observed at ~ 0.9, 1.9, and 2.35 eV in the $\varepsilon_1(\omega)$ for the ZIC, HIC and TIC phases, respectively. The $\varepsilon_1(\omega)$ becomes zero at around 10 eV for the TIC phase which corresponds to the energy at which the absorption coefficients vanishes (Fig. 6e), reflectivity exhibits a sharp drop (Fig. 6g) and the energy loss function (Fig. 6h) shows a first peak, however, similar behavior is not observed for the phases HIC and ZIC. The values of $\varepsilon_1(\omega)$ for ZIC, HIC and TIC phases go through zero at 2.5, 3.2 and 10 eV from below (Fig. 6a) while the values of $\varepsilon_2(\omega)$ (Fig. 6b) approach zero at 7.5, 20 and 22 eV from above, respectively, which assure that the studied phases are metallic in nature.

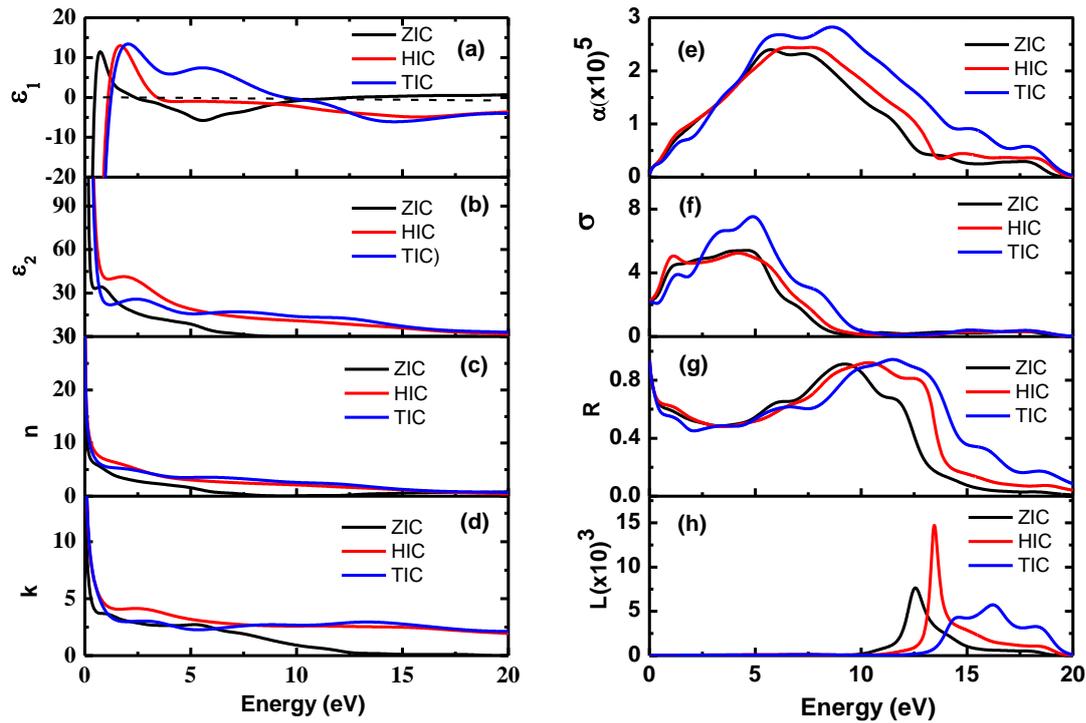

**Fig. 6.** Energy dependent (a) real part of dielectric function, (b) imaginary part of dielectric function, (c) refractive index, (d) extinction coefficient, (e) absorption coefficient, (f) photo conductivity, (g) reflectivity, (d) loss function of $Zr_2InC$ (ZIC), $Hf_2InC$ (HIC) and $Ta_2InC$ (TIC) phases for [100] polarization.

The refractive index *n* of crystals is a vital guide to design perfect electronic appliances. The frequency dependent behavior is shown in Fig. 6c. The absorption loss is defined by the extinction coefficient *k*. The calculated values of *k* of the phases are shown in Fig. 6d. The sharp peaks are obtained at 0- 2 eV for the phases causes intra-band transitions of electrons (Fig. 6d).

The values of absorption coefficient, *α* of the phases are shown in Fig. 6(e) which begins at 0 eV due to their metallic nature. Strong absorption coefficients are observed in the UV region for the phases, however it is weak in the IR region but continuously increases toward the UV region, and reaches a maximum value at 5.65, 7.85 and 8.70 eV for the ZIC, HIC and TIC, respectively. A material with high absorption coefficient is widely used in optical and optoelectronic devices in the visible and ultraviolet energy regions. Hence, this property makes the compounds ZIC, HIC and TIC promising candidates for optical and optoelectronic devices in the visible and ultraviolet energy regions. The photoconductivity starts at zero photon energy as shown in Fig. 6f as expected for metals. The TIC compounds show the highest photo-conductivity at around 5 eV photon energy whereas ZIC and HIC exhibit almost same photo-conductive behaviosr at near infrared, visible and near UV regions.

The reflectivity curves show that it starts with a value of 88% for the phases and rises to maximum values of 93% at 9.26 eV, 92% at 10.4 eV and 95% at 11.5 eV for the compounds ZIC, HIC and TIC, respectively (Fig. 6(g)). It is seen that the reflectivity spectra for all phases are almost constant in the visible light region and these values are always above 45% which is suitable for reducing solar heating in the visible light region [63]. Thus, the compounds are also promising candidates for the practical usage as a coating material to avoid solar heating. The reflectivity spectra for all phases approach zero at the incident photon energy range 19-22 eV.

The bulk plasma frequency $\omega_P$ can be obtained from the loss energy spectrum shown in Fig. 6(h). The energy loss function measures energy loss of an electron when it passes through a material. No loss spectra were found for any compounds in the energy up to 10 eV. The effective plasma frequency $\omega_P$ of the ZIC, HIC and TIC are found to be 12.6, 13.5 and 16.3 eV, respectively. The material becomes transparent, when the frequency of the incident light is higher than that of the plasma frequency.

## 4. Conclusions

We have presented a comparative study of the synthesized MAX phase ternary carbides $Zr_2InC$ and $Hf_2InC$ and predicted $Ta_2InC$ by employing the first-principles DFT calculations. Thermodynamic potentials, optical properties, electronic charge density distribution, Fermi surface topology, Mulliken bond overlap population and Vickers hardness have been investigated for the first time. The calculated elastic constants conform to mechanical stability conditions. The compounds are brittle in nature as expected for other MAX phases. The compounds are metallic in nature and the Zr-4$d$, Hf-5$d$ and Ta-5$d$ electrons are mainly contributing to the TDOS at the $E_F$. The strong covalent bond exists in the compounds and covalency level found in the chemical bonds of phases are articulated as C-Zr > In-Zr, C-Hf > In-Hf and C-Ta > In-Ta. The calculated Vickers hardness is found to be 1.05, 3.45 and 4.12 GPa for the phases ZIC, HIC and TIC, respectively that indicate the soft nature of studied phases. The dynamical stability of the compounds $M_2InC$ (M = Hf, Zr and Ta) has been confirmed using phonon dispersion curves. The optical properties of the compounds reveal several interesting properties of the phases. The reflectivity curves are always above 45% and rises to maximum values of 93% at 9.26 eV, 92% at 10.4 eV and 95% at 11.5 eV for the compounds ZIC, HIC and TIC, respectively. Therefore, the compounds are promising candidates for the optoelectronic device applications in the visible and ultraviolet energy regions and they can also be used as a coating material to avoid solar heating.

**Acknowledgements**

Authors are grateful to the Department of Physics, Chittagong University of Engineering & Technology (CUET), Chittagong-4349, Bangladesh, for arranging the financial support for this work.